\documentstyle[preprint,eqsecnum,aps]{revtex}
\tightenlines
\begin{document}
%
%
\title{
Fractional Kramers Equation
}
\author{
E. Barkai
\footnote{corresponding author. Fax: +1-617-253-7030; e-mail: barkai@mit.edu}
 and R.J. Silbey
\\Department of Chemistry\\and\\
Center for Materials Science and Engineering\\
M.I.T. Cambridge, MA}


\date{\today}
\maketitle

\begin{abstract}

 We introduce a fractional Kramers equation for a 
particle interacting with a thermal heat bath
and external non--linear force field. For the
force free case the velocity damping follows
the Mittag--Leffler relaxation and the
diffusion is enhanced.
The equation obeys the generalized Einstein relation,
and its stationary solution is the Boltzmann distribution.
Our results are compared to previous results
on enhanced  L\'evy type of diffusion 
derived from stochastic collision models.

\end{abstract}


%

\section{Introduction}

 The pioneering work of Scher and Montroll \cite{SM}
 and Scher and Lax \cite{SL} on the 
continuous time random walk \cite{Weiss1} applied to diffusion problems led to a revolution 
in our understanding of anomalous diffusion processes.  
Anomalous diffusion is now a well established phenomenon, 
found in a broad range
of fields \cite{Weiss1,Bouch,Klafter1,Balescu,barkai8}.  It is characterized by
a mean square displacement
\begin{equation}
\langle x^2 \rangle \sim t^{\delta}
\label{eqInt1}
\end{equation}
with $\delta \ne 1$. 
Various mechanisms are known to lead to
enhanced diffusion $\delta>1$ or sub-diffusion
$\delta<1$. Usually such processes are non-Gaussian
meaning that the standard central limit theorem cannot be used
to analyze the long time behavior of these phenomena.
In order to describe such anomalous processes, fractional kinetic equations 
were recently introduced by several authors
\cite{schneider,Gl,fogedby,zaslavsky,saichev,MBK,Hui,Grig,kuz}.
Within this approach,
fractional space and/or time derivatives replace
the ordinary time and/or space derivatives in the
standard kinetic equation (e.g., Fokker--Planck equation).
Examples include kinetics of viscoelastic media \cite{Gl},
L\'evy flights in random
environments \cite{fogedby}, chaotic Hamiltonian dynamics \cite{zaslavsky},
and Quantum L\'evy processes \cite{kuz}. 
For a discussion on L\'evy statistics and continuous time random walk
in the context of single molecule spectroscopy see \cite{Zum,SB,MUK}.

 In this paper, we introduce a fractional Kramers equation
describing both the velocity $v$ and coordinate $x$ of
a particle exhibiting anomalous diffusion in an external
force field $F(x)$.
In the absence of the external force field the equation describes
enhanced diffusion.
The new equation we propose is
$$ {\partial P(x,v,t) \over \partial t} + v {\partial P \over \partial x} + {F(x)\over M}
{\partial P(x,v,t) \over \partial v} = $$
\begin{equation}
\gamma_\alpha\ _0 D_t^{1-\alpha} \hat{L}_{fp} P(x,v,t),
\label{eqKr01a}
\end{equation}
with $0< \alpha < 1$ and
\begin{equation}
\hat{L}_{fp}=
{\partial \over \partial v} v + { k_b T \over M} { \partial^2 \over \partial  v^2} 
\label{eqR01a}
\end{equation}
 the dimensionless Fokker--Planck operator.
Here we employ  the fractional 
Liouville--Riemann operator \cite{oldham} $\ _0 D_t^{1-\alpha} $
in Eq. (\ref{eqKr01a}) which we define later  in
Eq. (\ref{eqL02}). 
$\gamma_{\alpha}$ is a damping coefficient whose units
are $[1/\mbox{sec}^{\alpha}]$.
In the presence of a bounding force
$F(x)=-V'(x)$ the stationary solution of the fractional 
Kramers equation (\ref{eqKr01a})
is the Maxwell--Boltzmann distribution. 
When $\alpha=1$ we recover the standard Kramers equation.

 Fractional kinetic equations are related to Montroll-Weiss continuous time
random walk (CTRW) \cite{Hilfer,compte,West,barkai9}. 
Here we show that the fractional Kramers equation is related
to the coupled CTRW in the enhanced diffusion regime $\delta > 1$.
This case corresponds to the so called L\'evy walks that are
observed in a number of systems[4,5].
The different limits of the CTRW were used to model
diverse physical phenomenon (when $F(x)=0$)
and therefore fractional kinetic equations in general
and the fractional Kramers equation in particular are believed
to be of physical significance.

 The basis of the fractional Kramers equation is the 
fractional Ornstein--Uhlenbeck process described by
the fractional Fokker--Planck equation
\begin{equation}
{\partial Q(v,t) \over \partial t} = 
\gamma_\alpha \ _0 D_t^{1-\alpha}\hat{L}_{fp}Q(v,t),
\label{eqR04aa}
\end{equation}
$Q(v,t)$  the probability density of finding the particle at
time $t$ with velocity $v$ when $F(x)=0$.
We see that the fractional Kramers equation is a natural
extension of Eq. (\ref{eqR04aa}) in which the streaming terms
describing Newtonian evolution are added in the standard way
(i.e., as in the Boltzmann or Liouville equations).
When $\alpha=1$ we get the standard Ornstein--Uhlenbeck process
which has a fundamental role in non equilibrium statistical mechanics.

 The Rayleigh model was used to derive Eq. (\ref{eqR04aa})
for the normal case of  $\alpha=1$ 
\cite{Ral,Kampen}.
The Rayleigh model for Brownian motion, also called the Rayleigh piston,
considers a one-dimensional heavy  particle 
of mass $M$ colliding with light non-interacting bath particles of mass $m$
which are always thermalized. 
According to this model the moments of  time intervals between collision events
are finite.
What happens when the variance of time intervals between collision events 
diverges? In this case we anticipate a non Gaussian behavior.
This case has been investigated by Barkai and Fleurov \cite{barkai3,barkai2}
and as we shall show in subsection 
\ref{secCM}, for a non-stationary model,
such an anomalous case corresponds to the
fractional Fokker--Planck equation (\ref{eqR04aa}). 
However, as we shall show, the usual Rayleigh limit $m/M \to 0$  
in the non-Gaussian case is not as straightforward as for the
ordinary Gaussian case.

 The more fundamental question of the necessary conditions for transport to
be described by diverging variance of time intervals between collision events,
or how to derive $\alpha$ from first principles is not addressed in this
paper.
In this context we note that
several mechanisms in which the variance of time between collisions
(or  turning events in random walk schemes) 
diverge and which lead to anomalous type of diffusion are known in the 
literature \cite{Bouch,Klafter1,barkai8}.

 Previous approaches 
\cite{schneider,Gl,fogedby,zaslavsky,saichev,MBK,Hui,Grig}
have considered fractional kinetic equations
in which the coordinate and/or time acquire the fractional character while in
our approach the velocities
are the variables that acquire fractional character. 
For $F(x)=0$ we find $1\le \delta=2-\alpha\le 2$.
The lower bound $\delta=1$ corresponds to normal diffusion, 
the upper bound $\delta=2$ can also be easily understood.
For a system close to thermal equilibrium we expect
\begin{equation}
\langle x^2 \rangle\le  (k_b T/M)t^2,
\label{bound}
\end{equation}
and hence $\delta\le 2$.

  Previously 
Kusnezov, Bulgac and Do Dang
\cite{kuz}
have suggested a fractional Kramers equation which was derived
for the classical limit of a Quantum L\'evy process. Also in this work do the 
velocities acquire fractional character; however this approach
is very different
than ours, since it is based on Reisz fractional operators and
 gives $\langle x^2(t) \rangle=\infty$ (i.e., $\delta=\infty$). 
Recently Metzler and Klafter \cite{MKp} considered a fractional kinetic equation
that they also called a fractional Kramers equation.  Their equation
describes sub-diffusion ($\delta < 1$) and is very different from our equation
which described enhanced diffusion ($\delta>1$).

 This paper is organized as follows.
In section \ref{sectwo} we introduce the fractional
Ornstein--Uhlenbeck process described by Eq. (\ref{eqR04aa}),
a brief introduction to fractional calculus is given.
In subsection
\ref{secsol}  the solution of the fractional Fokker--Planck 
Eq. (\ref{eqR04aa}) is presented  and in subsection
\ref{secCM} we derive this equation based on stochastic
collision model. In section 
\ref{Kra} we consider the fractional Kramers equation
(\ref{eqKr01a}), general properties of this equation are given
and the  force free case is investigated in some detail.
We end the paper with a short summary in section
\ref{secSum}.

\section{Fractional Ornstein--Uhlenbeck process}
\label{sectwo}

Let $Q(v,t)$ be the probability density describing the velocity
$v$ of a macroscopic Brownian particle, with mass $M$,
interacting with a thermal heat bath.
The Fokker--Planck equation describing the time evolution
of $Q(v,t)$ with initial conditions $Q(v,t=0)$
is given by \cite{Kampen,Risken}
\begin{equation}
{\partial Q\left( v, t \right) \over \partial t} = \gamma_1 \hat{L}_{fp} 
Q\left(v,t\right).
\label{eqR01}
\end{equation}
We shall call such a Fokker--Planck equation standard or
ordinary.
According to Eq. (\ref{eqR01})  the damping law 
for the averaged velocity is linear $\langle \dot{v}(t) \rangle=-
\gamma_1 \langle v(t) \rangle$, and the velocity fluctuations
are thermal. The stationary solution 
of Eq. (\ref{eqR01}) is the Maxwell's density
defined with the thermal energy
$k_b T$. 
The corresponding Langevin equation is 
\begin{equation}
\dot{v}=-\gamma_1 v + \xi(t)
\label{eqLang}
\end{equation}
and $\xi(t)$ is Gaussian white noise \cite{Kampen,Risken}. 
The stochastic process described by the Langevin equation is
the well known Ornstein--Uhlenbeck process.
Originally Eq. (\ref{eqR01})
was derived by Rayleigh \cite{Ral}
for a particle interacting with a gas consisting of
light particles, however the scope
of Eq. (\ref{eqR01}) is much wider. It is used to 
model Brownian motion in dense environments when memory effects
are negligible.


 In this section we generalize the Fokker--Planck equation (\ref{eqR01}) using fractional
calculus.
First we give some mathematical definitions and tools.

 The Liouville--Riemann fractional integral operator \cite{oldham}
of order $\alpha>0$ is defined by
\begin{equation}
_0 D_t^{-\alpha} f(t) \equiv \int_0^t{\left( t - t'\right)^{\alpha-1} \over \Gamma(\alpha)} f(t') d t'.
\label{eqL01}
\end{equation}
For integer values $\alpha=n$; $_0 D_t^{-n}$ is the Riemann
integral operator of order $n$. Fractional differentiation
of order $\alpha>0$ is defined by
\begin{equation}
_0 D_t^{\alpha} f(t) \equiv {d^n \over dt^n} \left[ _0 D^{\alpha-n}_t f\left(t\right)\right] ,
\label{eqL02}
\end{equation}
where $n-1\le \alpha < n$ and $d^n/dt^n$ is ordinary differentiation
of order $n$. Within this fractional calculus
\begin{equation}
_0 D^{\pm \alpha}_t t^{\mu} = { \Gamma\left( \mu+1 \right) \over \Gamma\left( \mu \mp \alpha +1 \right) } t^{\mu \mp \alpha}
\label{eqL03}
\end{equation}
when $\mu>-1$. Notice that $_0 D^{\pm \alpha}_t 1 \sim t^{\mp \alpha}$
when $0<\alpha<1$. The Laplace transform
\begin{equation}
f(u) = \int_0^{\infty} e^{ - u t } f(t) dt= {\cal L}[f(t)]
\label{eqL04}
\end{equation}
of a fractional Liouville--Riemann operator is
\begin{equation}
{\cal L} \left[ _0 D^{\alpha}_t f(t)\right] = u^{\alpha} f(u) - \sum_{k=0}^{n-1} u^{k}\  _0 D_t^{\alpha- 1 - k } f(t)|_{ t = 0}.
\label{eqL05}
\end{equation}
$n$ is an integer satisfying $n-1< \alpha \le n$.
For fractional integrals $\alpha\le 0$
the sum on the RHS of Eq. (\ref{eqL05})
vanishes. From  Eq. (\ref{eqL05}) 
we see that the Laplace transform of a fractional
derivative of $f(t)$ depends on fractional derivatives
of that function at time $t=0$.
In this work we use the convention that the arguments
of a function indicate the space in which the function
is defined, e.g. the Laplace transform of $Q(v,t)$ is
$Q(v,u)$.

 The Fokker--Planck equation (\ref{eqR01})
is rewritten in the integral form
\begin{equation}
Q(v,t) - Q(v,t=0)= \gamma_1\ _0 D_t^{-1}\hat{L}_{fp} Q(v,t)
\label{eqR02}
\end{equation}
We now replace the integer integral operator $_0 D_t^{-1}$ 
in Eq. (\ref{eqR02}) with a fractional integral operator
$_0 D_t^{-\alpha}$ and $0<\alpha\le 1$. The result
is
\begin{equation}
Q(v,t) - Q(v,t=0)= \gamma_\alpha \ _0 D_t^{-\alpha}\hat{L}_{fp} Q(v,t)
\label{eqR03}
\end{equation}
where $\gamma_{\alpha}$ is a generalized damping coefficient
with units $[\mbox{sec}]^{-\alpha}$.
Ordinary differentiation of Eq. (\ref{eqR03})
gives the fractional Fokker--Planck equation
\begin{equation}
{\partial Q(v,t) \over \partial t} = \gamma_\alpha \ _0 D_t^{1-\alpha}\left[{\partial \over \partial v} v + { k_b T \over M} { \partial^2 \over \partial v^2} \right] Q(v,t).
\label{eqR04}
\end{equation}
When $\alpha=1$ the ordinary Fokker--Planck  Eq.
(\ref{eqR01})
is obtained.
In Eq. (\ref{eqR04}) we use the  natural boundary conditions
\begin{equation}
\lim_{v\to \pm \infty} Q(v,t)=\lim_{v\to  \pm \infty} \partial Q(v,t)/\partial v=0.
\label{eqR04a}
\end{equation}
Later we shall show that the solution of Eq. (\ref{eqR04})
is non-negative and normalized. Eq. (\ref{eqR04}) describes a fractional
Ornstein--Uhlenbeck process. When the velocity $v$ is replaced 
with a coordinate
$x$, the equation describes an anomalously over damped
 harmonic oscillator as investigated in \cite{MBK}.

 Eqs. (\ref{eqR03}) and (\ref{eqR04}) are initial
value problems.
While Eq. (\ref{eqR03}) depends on a single initial condition
[$Q(v,t=0)$ on the LHS of the equation],
in solving Eq. (\ref{eqR04}) two initial conditions
have to be specified \cite{oldham},
these being $Q(v,t=0)$ and $_0D_t^{-\alpha}Q(v,t)|_{t=0}$.
When setting $_0D_t^{-\alpha}Q(v,t)|_{t=0}$ equal to zero, the two
equations are equivalent
\cite{remark3}.

 Multiplying Eq. (\ref{eqR04}) by $v$ and integrating over 
$v$ we find
that the  mean velocity is described by
\begin{equation}
\langle \dot{v}\left( t \right) \rangle= 
- \gamma_{\alpha}\ _0 D^{ 1 - \alpha}_t \langle v\left( t \right) \rangle.\label{eqR07}
\end{equation}
In Laplace space Eq. (\ref{eqR07}) reads
\begin{equation}
\langle v(u) \rangle ={ v_0 \over u + \gamma_{\alpha} u^{ 1 - \alpha}}.
\label{eqR08}
\end{equation}
$v_0$ is the initial velocity.
The inverse Laplace transform of Eq. (\ref{eqR08}) is
\begin{equation}
\langle v(t) \rangle = v_0 E_{\alpha}\left( - \gamma_{\alpha} t^{\alpha} \right),
\label{eqR09}
\end{equation}
and 
\begin{equation}
E_{\alpha}\left( z \right) =\sum_{n = 0 }^{\infty} { z^n \over \Gamma\left( 1 + \alpha n \right)}
\label{eqR10}
\end{equation}
is the Mittag--Leffler function.
When $\alpha=1$ the Mittag--Leffler reduces to the
exponential. For large $t$, 
Eq. (\ref{eqR09}) exhibits a power law decay
\begin{equation}
\langle v(t) \rangle \sim {v_0 \left( \gamma_{\alpha} t\right)^{- \alpha} \over \Gamma\left( 1 - \alpha\right) } 
\label{eqvvv}
\end{equation}
and for short times the relaxation is a stretched exponential
\begin{equation}
\langle v(t ) \rangle \simeq v_0 \exp\left[ - { \gamma_{\alpha} t^{\alpha} \over \Gamma\left( 1 + \alpha\right) } \right].
\label{eqsstt}
\end{equation}
In a similar way we find the second moment
\begin{equation}
\langle v^2(t) \rangle = v_0^2 E_{\alpha}\left( - 2 \gamma_{\alpha} t^{\alpha}\right)+ { k_b T \over M} \left[ 1 - E_{\alpha} \left( - 2 \gamma_{\alpha} t^{\alpha} \right)\right],
\label{eqv22}
\end{equation}
exhibiting power law decay towards the thermal equilibrium value
$\langle v^2 \left(t=\infty\right)\rangle =k_b T /M$.
From Eqs. (\ref{eqR09}) and
(\ref{eqv22}) we see that the Mittag--Leffler relaxation
replaces the ordinary exponential relaxation found for ordinary Brownian motion.
These equations were  derived in \cite{barkai3}  based upon a stochastic collision model
which we will discuss in subsection \ref{secCM}.

\subsection{Solution}
\label{secsol}

 Our aims are {\bf (a)} to find a solution of the fractional Fokker--Planck
Eq. (\ref{eqR04}) and {\bf (b)} show that $Q(v,t)$ in Eq.
(\ref{eqR04}) is a probability density. 
The solution we find is an integral of a product of two
well known functions.
We use the initial conditions
$Q(v,t=0)= \delta(v - v_0)$ the generalization for other
initial conditions is carried out in the usual way.

 We first show that the solution is normalized.
The Laplace transform of Eq. 
(\ref{eqR04}) is
\begin{equation}
uQ(v,u)-\delta(v-v_0)=\gamma_{\alpha}u^{1 - \alpha}\hat{L}_{fp} Q(v,u).
\label{eqA01}
\end{equation}
Integrating Eq. (\ref{eqA01}) with respect to $v$,
using the boundary conditions
in Eq. (\ref{eqR04a}) and the normalized initial condition
we find
\begin{equation}
\int_{-\infty}^{\infty} Q(v,u) dv= {1 \over u}.
\label{eqANOR}
\end{equation}
Since ${\cal L}(1,u) = 1/u$ we see that $Q(v,t)$ in 
Eq. (\ref{eqR04}) is normalized.

 Let us now find the solution in Laplace $u$ space.
We write $Q(v,u)$ as
\begin{equation}
Q(v,u)=\int_0^{\infty} R_{s}(u) G_{s}( v ) d s
\label{eqA04}
\end{equation}
where
\begin{equation}
\hat{L}_{fp} G_{s} \left( v \right) = { \partial \over \partial s} G_{s} \left( v \right)
\label{eqA05}
\end{equation}
and
\begin{equation}
 G_0\left( v \right) = \delta( v - v_0).
\label{eqA05ic}
\end{equation}
Eq. (\ref{eqA05}) is the dimensionless ordinary 
Fokker--Planck Eq. (\ref{eqR01}),
with solution \cite{Kampen,Risken}
$$ G_{s} \left( v \right) = $$
\begin{equation}
{ \sqrt{M} \over \sqrt{ 2 \pi k_b T \left( 1 - e^{ - 2 s} \right)}}
\exp\left[ - { M\left( v - v_0e^{ - s} \right)^2 \over
2 k_b T \left( 1 - e^{ - 2 s} \right) } \right].
\label{eqA06m}
\end{equation}
We see that $G_s(v)$ is a non-negative probability density function,
describing the standard 
 Ornstein--Uhlenbeck process, and normalized according to
\begin{equation}
\int_{-\infty}^{\infty} G_s(v) dv = 1.
\label{eqadd}
\end{equation}

 $R_{s}\left( u \right)$ in Eq.
(\ref{eqA04}) must satisfy a normalization condition.  Using  Eqs. 
(\ref{eqANOR}) and (\ref{eqadd}), we have
\begin{equation}
\int_0^{\infty} R_{s} \left( u \right)d s= {1 \over  u}.
\label{eqA07}
\end{equation}
Inserting Eq. (\ref{eqA04}) in Eq. (\ref{eqA01}), using
Eq. (\ref{eqA05}), and integrating by parts, we find 
$$ u\int_0^{\infty} R_{s} \left( u \right) G_{s} ( v ) d s- \delta(v-v_0) = $$
$$ \gamma_{\alpha} u^{1 - \alpha} \int_0^{\infty} R_{s} \left( u \right) \hat{L}_{fp} G_{s}\left( v \right) d s =$$
$$ \gamma_{\alpha} u^{1 - \alpha} \int_0^{\infty} R_{s} \left( u \right) 
{ \partial \over \partial s} G_{s}\left( v \right) d s 
= $$
$$ \gamma_{\alpha} u^{1 - \alpha} \left[ R_{\infty}\left( u \right) 
G_{\infty}\left( v \right) - R_0\left( u \right) G_0\left( v \right) \right]-$$
\begin{equation}
\gamma_{\alpha} u^{ 1 - \alpha} \int_0^{\infty}
 \left[ {\partial \over \partial s} R_{s}\left( u \right) \right] G_{s}\left( v \right) d s.
\label{eqA08}
\end{equation}
According to Eq. (\ref{eqA07}) the  boundary term 
$R_{\infty}(u)$
in
Eq. (\ref{eqA08})  is zero.
Using the initial condition Eq.
(\ref{eqA05ic})
 in Eq. 
(\ref{eqA08})
we find
$$ \int_0^{\infty}\left\{ u R_{s}\left(u\right)+\gamma_{\alpha} u^{1-\alpha}\left[
{\partial \over \partial
s} R_{s}\left( u \right) \right] \right\} G_{s}\left( v \right) d s=$$
\begin{equation}
\left[1 - \gamma_{\alpha}u^{1 - \alpha} R_0(s)\right]\delta(v - v_0).
\label{eqexp}
\end{equation}
Eq. (\ref{eqexp}) is solved once both sides of it are equal
to zero; therefore, two conditions must be satisfied,  the first being
\begin{equation}
\gamma_{\alpha} u^{1 - \alpha} R_0 \left( u \right) = 1
\label{eqA09}
\end{equation}
and the second being
\begin{equation}
- \gamma_{\alpha} u^{ 1 - \alpha} { \partial \over \partial s} R_{s} \left( u \right) = u R_{s} \left( u \right).
\label{eqA10}
\end{equation}
The solution of Eq. (\ref{eqA10}) with the condition Eq. (\ref{eqA09})
is
\begin{equation}
R_{s}\left( u \right) = { 1 \over \gamma_{\alpha} u^{ 1 - \alpha} } \exp\left( 
- { s u^{\alpha} \over \gamma_{\alpha} } \right).
\label{eqA11m}
\end{equation}
It is easy to check that $R_{s}\left( u \right)$ is normalized according
to Eq. 
(\ref{eqA07}).

 The solution of the problem in $t$ space is the inverse Laplace of Eq.
(\ref{eqA04})
\begin{equation}
Q(v,t)=\int_0^{\infty} R_{s}(t) G_{s}( v ) d s,
\label{eqA04mt}
\end{equation}
where $R_{s}(t)$ is the inverse Laplace transform of $R_{s}(u)$ given by
\begin{equation}
R_{s}\left( t \right) = { 1 \over \alpha \gamma_{\alpha} t^{\alpha}}
z^{\alpha + 1} l_{\alpha}\left( z \right),
\label{eqlevy}
\end{equation}
and 
\begin{equation}
z={ \left( \gamma_{\alpha}\right)^{1/\alpha} t \over s^{1/\alpha} }.
\label{eqlevyz}
\end{equation}
Properties of $R_s(t)$ are discussed by Saichev and Zaslavsky \cite{saichev}.
$l_{\alpha}\left( z \right)$ in Eq. (\ref{eqlevy}) is one sided L\'evy stable
density  
\cite{Feller}, 
whose Laplace transform is 
\begin{equation}
l_{\alpha}(u)=\int_0^{\infty} \exp\left( - u z\right) l_{\alpha}\left( z \right) dz =\exp( -u^{\alpha}).
\label{eqadd1}
\end{equation}
The proof that $R_{s}( t )$ Eq. (\ref{eqlevy}) and 
$R_{s}( u )$ Eq.(\ref{eqA11m}) are a Laplace pair is given in 
Appendix A (and see also \cite{saichev}).

 A few features of the solution Eq. (\ref{eqA04mt}) can now be discussed.
An interpretation of Eq.(\ref{eqA04mt}) in terms of a stochastic 
collision model will be given in the next subsection.

 When $0< \alpha \le 1$; $R_{s}(t)$ is a probability density normalized
according to $\int_0^{\infty} d s R_{s}(t)=1$.
Since  $G_{s}(v)$ is also a probability density
the solution Eq. (\ref{eqA04mt}) is normalized and non negative.
This justifies our interpretation of $Q(v,t)$ 
as a probability density. 

 When $\alpha=1$ the solution reduces to the well known
solution of the ordinary Fokker--Planck equation. 
To see this, note that the inverse Laplace of Eq.
(\ref{eqA11m}) for $\alpha=1$  is
$R_{s}(t) = {1/\gamma_1} \delta\left( t -s/\gamma_1 \right)$,
and then use definition Eq.
(\ref{eqA04mt}).
 When $\alpha=1/2$ we have 
\begin{equation}
l_{1/2}(z)={1 \over 2 \sqrt{\pi}}  z^{-3/2} \exp\left( - {1 \over 4 z}\right)
\label{eqlevyha}
\end{equation}
with $z>0$, then $R_{s}(t)$ 
Eq. (\ref{eqlevy})
is a one sided
Gaussian
\begin{equation}
R_{s}(t)=\sqrt{{ 1 \over \pi \gamma_{1/2}^2 t}} \exp\left( - { s^2 \over 4 \gamma_{1/2}^2 t } \right).
\label{eqpthalf}
\end{equation}
Two other closed forms of one sided L\'evy probability
densities $l_{2/3}(z)$ and $l_{1/3}(z)$ can be found
in \cite{Zol,EWM}. Series expansions of L\'evy
stable density are in Feller's book \cite{Feller} chapter
{\bf XVII.6}. 

 In Fig. 1, we show the solution 
for the case $\alpha = 1/2$ and for different
times. The solution is found  with numerical integration
of 
(\ref{eqA04mt}) using Eqs. 
(\ref{eqA06m}) and
(\ref{eqpthalf}).  We choose the initial condition
$Q(v,t=0)=\delta(v-1)$, (i.e., $v_0 = 1$) and 
$k_b T / M = 1$. The solution exhibits a slow power law  decay towards 
thermal equilibrium.  We observe a cusp at $v=v_0=1$; thus, initial conditions
have a strong signature on the shape of $Q(v,t)$. 
A close look at the figure shows that for short times (i.e., $t\le 2$) the peak of
$Q(v,t)$ is at $v=v_0$. This is very different from Gaussian evolution
for which the peak is always on $\langle v(t) \rangle$.





 Metzler, Barkai and Klafter \cite{MBK}
have shown that a fractional Fokker--Planck equation,
which describes sub-diffusion $\delta<1$,
can be solved using an eigenfunction expansion which is identical 
to the ordinary expansion of the Fokker--Planck solution \cite{Risken};
but in which the  exponential relaxation of eigenmodes is replaced with a
Mittag--Leffler relaxation.
We can use the eigenfunction expansion in \cite{MBK}
to find a second  representation of $Q(v,t)$  
in terms of a sum of Hermite polynomials. Expanding $G_{s}\left( v \right)$,
using the standard eigenfunction technique of Fokker--Planck
solutions, we write
$$G_s\left( v \right) = \sqrt{{ M \over 2 \pi k_b T}} \exp\left( - { M v^2 \over 2 k_b T } \right) \times $$
\begin{equation}
\sum_{n = 0}^{\infty} { 1 \over 2^n n! } H_n\left(\sqrt{{M \over 2 k_b T}} v \right)
H_n\left(\sqrt{{M \over 2 k_b T}} v_0 \right) \exp\left(-n s\right)
\label{eqexp1}
\end{equation}
where $H_n$ are Hermite polynomials. We insert the expansion
Eq. (\ref{eqexp1}) in Eq.
(\ref{eqA04mt}) and use
\begin{equation}
\int_0^{\infty} R_{s}\left( t \right) \exp\left( - n s\right) d s=
E_{\alpha}\left( - n \gamma_{\alpha} t^{\alpha} \right),
\label{eqidentity}
\end{equation}
to find the eigen function expansion 
$$Q \left(v, t \right) = \sqrt{{ M \over 2 \pi k_b T}} \exp\left( - { M v^2 \over 2 k_b T } \right) \times $$
\begin{equation}
\sum_{n = 0}^{\infty} { 1 \over 2^n n! } H_n\left(\sqrt{{M \over 2 k_b T}} v \right)
H_n\left(\sqrt{{M \over 2 k_b T}} v_0 \right) E_{\alpha} \left(-n \gamma_{\alpha} t^{\alpha} \right).
\label{eqexp2}
\end{equation}
The stationary solution, determined by the smallest eigen value $n=0$,
is the Maxwell distribution which is independent of $\gamma_{\alpha}$
and $\alpha$.

 An extension of the fractional Fokker--Planck equation
(\ref{eqR04}) to higher dimensions is carried out by replacing the
one-dimensional Fokker--Planck operator, Eq. 
(\ref{eqKr01a})
 with the appropriate
$d$ dimensional Fokker--Planck operator (e.g., replace ${\partial / \partial v}$
with $\nabla$). The solution for such a $d$ dimension equation
is then found to be Eq. 
(\ref{eqA04mt}) in which  Eq.
(\ref{eqA06m}) must be replaced with the appropriate solution
of the $d$ dimensional ordinary Fokker--Planck equation.

\subsection{Collision Model}
\label{secCM}

 As mentioned above, the ordinary Fokker--Planck equation
(\ref{eqR01}) was derived by
Rayleigh over a century ago. Briefly, the Rayleigh model
for Brownian motion considers a one dimensional test particle with mass $M$
colliding with bath particles of mass $m$ and $\epsilon\equiv m/M<<1$.
When collisions are frequent but weak,  the ordinary Fokker--Planck
equation is valid \cite{Kampen}. 
Here we consider the case when  the concept of
collision rate does not hold and 
the mean time
intervals between collision events diverges.
As mentioned in the introduction this case was investigated in \cite{barkai3,barkai2}
and as we shall show now such a case corresponds to the
fractional Fokker--Planck equation 
(\ref{eqR04}).

 We consider a particle of mass $M$ which moves freely in
one dimension and at random times it collides elastically
with bath particles of mass $m$.
Bath particles are assumed to be much faster than the test particle. 
Collisions are elastic and one-dimensional and therefore the velocity
$V_M^+$ of the test particle immediately after a
collision event can be related to the velocity $V_M^-$ of the
test particle just before the collision event  
according to
\begin{equation}
V_M^+ = \left({ 1- \epsilon \over 1 + \epsilon}\right) V_M^-
 + {2 \epsilon \over 1 + \epsilon} \tilde{v}_m.
\label{eqCol01}
\end{equation}
where ${\epsilon} = m/M$ and $\tilde{v}_m$ is the velocity of bath particle
distributed according to Maxwell's distribution.

 The times between collision events are assumed to be independent
identically distributed random variables implying
that the number of collisions in a time interval
$(0,t)$ is a renewal process. This is reasonable
when the bath particles thermalize very quickly and when
the test particle is slow. According to these assumptions the
times between collision events  $\{ \tau_i \}$ are described by a probability
density $\psi\left( \tau \right)$ which is independent 
of the mechanical state of the test particle.  Therefore the process
is characterized by free motion with constant velocity for
time $\tau_1$ then a collision event described by Eq. (\ref{eqCol01}) 
and then a free evolution for a period $\tau_2$, then again a collision etc.
The most important ingredient of the model is the assumption
that $\psi(\tau)$ decays like a power law for long time
\begin{equation}
\psi(\tau) \sim \tau^{ - 1 - \alpha}
\label{eqCol02}
\end{equation}
with $0<\alpha <1$. From Eq. (\ref{eqCol02}) we learn that the mean time
between collision diverges, $\int^{\infty}_0 \tau \psi(\tau) d\tau=\infty$.
Since in such a problem there is no characteristic time scale,  the number
of collisions in an interval $(0,t)$ is not proportional to $t$ for large
times. In other words the law of large numbers is not valid for the 
choice  Eq. (\ref{eqCol02}), leading to non normal behavior.
Similar waiting times distributions were used within
the CTRW to model anomalous diffusion for the past three decades.
When $\psi(\tau)$ is exponential and in the presence of an external force field $F(x)$,
this model was investigated extensively in the context of
reaction rate theory \cite{Hangi,Mont,Knessel,Bork,barkaif,Ber1,Ber2}.

 Such a model is non stationary and the probability of a collision
event in a small time interval $(t,t+dt)$ is time dependent
even in the limit of large times. This is a consequence of the diverging
first moment.

 Let $Q_{col}\left( v, t \right)$ be the probability density for
finding the test particle with velocity $v$ at time $t$ and initially
$v=v_0$. Using the model assumptions
\begin{equation}
Q_{col}\left( v , t \right) = \sum_{s=0}^{\infty} \tilde{R}_s\left( t \right) \tilde{G}_s \left( v \right)
\label{eqCol03}
\end{equation}
with $\tilde{R}_s(t)$ is the probability that $s$ collision
events have occurred in the interval $(0,t)$
and $\tilde{G}_s\left( v \right)$ is the conditional probability 
density of finding the particle with velocity $v$ after $s$ collision
events. We note that Eq. (\ref{eqCol03}) is the discrete version
of Eq. 
(\ref{eqA04mt}).
 
Using the map Eq. (\ref{eqCol01}), it can be shown that
\begin{equation}
\tilde{G}_s\left( v \right) = { \sqrt{M} \over \sqrt{ 2 \pi k_b T \left( 1 - \mu_1^{2s} \right)}}
\exp\left[ - { M \left( v - v_0\mu_1^s\right)^2 \over 2 k_b T \left( 1 - \mu_1^{2 s} \right)} \right]
\label{eqCol04}
\end{equation}
with $\mu=(1-\epsilon)/ ( 1 + \epsilon)$.
Not surprisingly $\tilde{G}_s(v)$ is Gaussian since velocities of colliding
particles are Gaussian random variables. We note that
\begin{equation}
{\partial \tilde{G}_s\left( v \right) \over \partial s} =
 \ln\left(\mu_1^{-1} \right)\hat{L}_{fp}\tilde{G}_s\left( v \right) 
\label{eqCol05}
\end{equation}
with the initial condition
\begin{equation}
\tilde{G}_0\left( v \right) = \delta\left( v - v_0\right)
\label{eqCol05a}
\end{equation}
and $\ln\left( \mu_1^{-1} \right) \sim 2 \epsilon$. Eq. (\ref{eqCol05}) is 
a Fokker--Planck equation in which $s$, the collision number, plays the role of dimensionless
time.

 The Laplace $t \to u$ transform of $\tilde{R}_s\left( t \right)$, 
$\tilde{R}_s\left( u \right)$
can be calculated using renewal theory
\begin{equation}
\tilde{R}_s\left( u \right)= { 1 - \psi(u) \over u} \psi^s(u)
\label{eqCol06}
\end{equation}
and $\psi(u)= \cal{L}[\psi(\tau)]$. From Eq.
(\ref{eqCol03}) we have in Laplace $t \to u$ space
\begin{equation}
Q_{col}\left( v , u \right) = \sum_{s=0}^{\infty} \tilde{R}_s\left( u \right) \tilde{G}_s\left( v \right),
\label{eqCol07}
\end{equation}
multiplying this equation from the left with
$\hat{L}_{fp}$,  using Eq.(\ref{eqCol05}), and integrating by parts, we have
$$\hat{L}_{fp} Q_{col}\left( v , u \right) ={1 \over 2 \epsilon}  \sum_{s=0}^{\infty} \tilde{R}_s\left( u \right) 
{\partial  \tilde{G}_s\left( v \right) \over \partial s}$$
\begin{equation}
= -{ 1 \over 2 \epsilon} \tilde{R}_0\left( u \right) \delta \left( v- v_0 \right) 
- { 1 \over 2 \epsilon} \sum_{s = 0 } ^\infty 
\left[ {\partial \over \partial s} \tilde{R}_s 
\left( u \right) \right] \tilde{G}_s \left( v \right),
\label{eqCol09}
\end{equation}
We have used Eq.
(\ref{eqCol05a}) and the the boundary condition 
$\tilde{R}_{\infty}\left( u \right) = 0$ for $u \ne 0$.

 According to Eq. 
(\ref{eqCol06}) 
\begin{equation}
{\partial \over \partial s}  \tilde{R}_s\left( u \right)= 
\ln \left[ \psi( u ) \right] \tilde{R}_s\left( u \right)
\label{eqCol10}
\end{equation}
and since according to
(\ref{eqCol02})
$\psi(\tau) \sim \tau^{ - 1 - \alpha}$ we have
\begin{equation}
\psi\left( u \right) \sim 1 - A u^{\alpha}
\label{eqCol11}
\end{equation}
(here A is a parameter with units $t^{\alpha}$)
valid for small $u$, inserting Eq. (\ref{eqCol11}) in Eq. (\ref{eqCol10})
we have
\begin{equation}
{\partial \over \partial s} \tilde{R}_s\left( u \right)\sim  - Au^{\alpha } R_s\left( u \right)
\label{eqCol12}
\end{equation}
and from Eq.
(\ref{eqCol06}) 
\begin{equation}
\tilde{R}_0(u)\sim Au^{\alpha - 1}.
\label{eqadd2}
\end{equation}

We are now ready to derive the fractional Fokker--Planck equation.
Inserting Eqs. (\ref{eqCol12}) and
(\ref{eqadd2})
 in 
Eq. (\ref{eqCol09}) 
we find in the limit of small $A u^{\alpha} $ and $\epsilon$
\begin{equation}
uQ_{col}\left( v, u \right) - \delta\left( v - v_0\right)= 
\gamma_{\alpha}u^{1 - \alpha} \hat{L}_{fp} Q_{col}\left( v , u \right)
\label{eqColl13}
\end{equation}
with
\begin{equation}
\gamma_{\alpha} = \lim_{A\to 0, \epsilon \to 0} { 2 \epsilon \over A}.
\label{eqColl14}
\end{equation}
Eq. (\ref{eqColl13}) is the fractional Fokker--Planck equation 
(\ref{eqR01})
in Laplace
space.
  We note that the moments of the collision model
in the limit $\epsilon \to 0$ found in \cite{barkai3} 
are identical to those found here based upon the fractional Fokker--Planck equation
Eqs. 
(\ref{eqR09}) and
(\ref{eqv22}) as they should be.

\section{Fractional Kramers Equation}
\label{Kra}

 Let $P(x,v,t)$ be the joint probability density function
describing both the position $x$ and the velocity $v$ of a Brownian
particle subjected to an external force field $F(x)$.
The one dimensional Kramers
equation models such stochastic motion according to
$$ {\partial P(x,v,t) \over \partial t} + v {\partial P \over \partial x} + {F(x)\over M}
{\partial P(x,v,t) \over \partial v} = $$
\begin{equation}
\gamma_1 \hat{L}_{fp} P(x,v,t),
\label{eqKr01}
\end{equation}
where the Fokker--Planck operator $\hat{L}_{fp}$ 
is given in Eq. 
(\ref{eqR01a}). 
The Kramers Eq. (\ref{eqKr01}) implies that noise is white
and Gaussian and it describes under--damped motion close
to thermal equilibrium. 
Eq. (\ref{eqKr01}) 
is an extension of Eq. (\ref{eqR01}) which includes the coordinate
$x$ as well as  the effects
of $F(x)$. We generalize Kramers equation in the same way as above,
 and consider
$$ {\partial P(x,v,t) \over \partial t} + v {\partial P(x,v,t) \over \partial x} + {F(x)\over M} 
{\partial P(x,v,t) \over \partial v} = $$
\begin{equation}
\gamma_\alpha\ _0 D_t^{1-\alpha} \hat{L}_{fp} P(x,v,t),
\label{eqKr02}
\end{equation}
with $0<\alpha<1$.
The terms on the LHS of the equation
are the standard streaming terms describing
reversible dynamics according to Newton's second law of motion.
The term
on the RHS of the equation describes an interaction
with a bath, it can be considered as a generalized collision
operator replacing the ordinary collision operator found
in the standard Fokker--Planck equation. 
As mentioned in the introduction the stationary 
solution of Eq. (\ref{eqKr02}) is the 
Maxwell--Boltzmann distribution and
when $\alpha=1$ we recover the ordinary Kramers equation.
 
 A formal solution of the fractional Kramers equation can be found in
terms of the solution of the ordinary Kramers equation.
We denote the solution of the fractional Kramers equation
with $P_{\alpha} \left( x,v,t,\gamma_{\alpha}\right)$ instead of
$P(x,v,t)$ we have used so far. 
The Laplace transform 
of Eq. (\ref{eqKr02}) is 
$$u P_{\alpha}(x,v,u,\gamma_{\alpha}) -P_{\alpha}\left( x,v,t=0,\gamma_{\alpha}\right)$$
$$ + v {\partial P_{\alpha}\left(x,v,u,\gamma_{\alpha}\right) \over \partial x} + {F(x)\over M} 
{\partial P_{\alpha}(x,v,u,\gamma_{\alpha}) \over \partial v} = $$
\begin{equation}
\gamma_\alpha\ u^{1-\alpha} \hat{L}_{fp} P_{\alpha}(x,v,u,\gamma_{\alpha}),
\label{eqKr02a}
\end{equation}
and $P_{\alpha}\left(x,v,u,\gamma_{\alpha}\right)$ 
is the Laplace transform of $P_{\alpha}\left(x,v,t,\gamma_{\alpha}\right)$. 
From Eq. (\ref{eqKr02a}) we learn that 
$P_{\alpha}(x,v,u,\gamma_{\alpha})$ solves an ordinary Kramers equation 
in which the damping coefficient $\gamma_1$ was transformed according to
\begin{equation}
\gamma_1 \to \gamma_{\alpha} u^{1 - \alpha}.
\label{eqwww}
\end{equation}
We therefore find
\begin{equation}
P_{\alpha} \left( x,v , u , \gamma_{\alpha}\right) = 
P_1\left( x, v, u, \gamma_{\alpha} u^{1 - \alpha}\right),
\label{eqwww11}
\end{equation}
assuming the initial conditions are identical
for both solutions. Transforming to the time domain we find the formal solution
\begin{equation}
P_{\alpha}\left( x, v, t, \gamma_{\alpha} \right) =
{\cal L}^{-1} \left[ P_1\left( x, v, u, \gamma_{\alpha}u^{1 - \alpha} \right) \right]
\label{eqwww2}
\end{equation}
with ${\cal L}^{-1}$ being the inverse Laplace transform.
Closed form solutions of ordinary Kramers equation, $P_1\left( x, v, t, \gamma_1\right)$,
are known only for a handful of cases, approximate solutions can be found
using methods specified in \cite{Risken}.
In some cases
Eq. (\ref{eqwww2}) can be used to find moments of the solution of fractional
Kramers equation, $\langle x^n v^m \rangle$, in a straight forward way.
In what follows we shall revert to the 
notation $P(x,v,t)$ instead of $P_{\alpha}(x,v,t,\gamma_{\alpha})$.

\subsection{Force free case 1}

  We consider the force free case $F(x)=0$.
As usual $\langle \dot{x}(t) \rangle = \langle v(t) \rangle$,
with the mean velocity $\langle v(t) \rangle$ given in Eq.
(\ref{eqR09}), hence
\begin{equation}
\langle x(t) \rangle = v_0 t E_{\alpha,2} \left( - \gamma_{\alpha} t^{\alpha} \right)\label{eqKr03}
\end{equation}
where
\begin{equation}
E_{\alpha,\beta}(z)=\sum_{k=0}^{\infty} { z^k \over \Gamma\left( \alpha + \beta k \right)}
\label{eqKr04}
\end{equation}
is a generalized 
Mittag--Leffler function \cite{Erde} satisfying
\begin{equation}
E_{\alpha,\beta}\left( z \right) = - \sum_{n=1}^{N-1} { z^{-n} \over \Gamma\left( \beta - \alpha n\right) } + {\cal O} \left( z^{-n}\right)
\label{eqKr05}
\end{equation}
with $z \to \infty$. For short times
\begin{equation}
\langle x(t) \rangle \sim v_0 t
\label{eqKr05a}
\end{equation}
as expected for a pure ballistic
propagation, and for long times
\begin{equation}
\langle x(t) \rangle \sim {v_0 t^{1-\alpha}\over \gamma_{\alpha} \Gamma\left( 2 - \alpha\right)}.
\label{eqKr0r65}
\end{equation}
The particle exhibits a net drift in a direction determined
by the initial velocity $v_0$. Of course,
when averaging over initial conditions, using
thermal equilibrium condition, no net drift is
observed as expected from symmetry. The mean
square displacement is determined by $\langle \dot{x}^2(t)\rangle=
\langle 2 x(t) v(t) \rangle$. A short calculation using the Laplace transform
 of Eq. 
(\ref{eqKr02}) shows
\begin{equation}
\langle x^2(t) \rangle_{eq} = 2 {k_b T \over M} t^2 E_{\alpha, 3}\left( - \gamma_{\alpha}t^{\alpha} \right), 
\label{eqKr07}
\end{equation}
where the subscript $_{eq}$ means that thermal initial
conditions are considered (i.e., $\langle v_0^2 \rangle_{eq}=k_b T / M$).
For short times
\begin{equation}
\langle x^2(t) \rangle_{eq} \sim  {k_b T \over M} t^2, 
\label{eqKr08}
\end{equation}
while for long times
\begin{equation}
\langle x^2(t) \rangle_{eq} \sim  2 D_{\alpha} t^{ 2 - \alpha} 
\label{eqKr09}
\end{equation}
where
\begin{equation}
 D_{\alpha} = { k_b T \over \gamma_{\alpha} M \Gamma\left( 3 - \alpha\right) }.
\label{eqKr10}
\end{equation}
Eq. (\ref{eqKr09}) exhibits an enhanced diffusion when $0<\alpha<1$.
Eq. (\ref{eqKr10}) is the (first) generalized Einstein relation  and
when $\alpha=1$ we recover the well known Einstein relation
$D_1=(k_b T ) /(M \gamma_1)$.

It is straightforward to prove the more general Einstein relation 
\begin{equation}
\langle x^2 \left( t \right) \rangle_{eq} = 2 \int_0^t dt' \int_0^{t'} \langle 
v(\tau) v(0) \rangle_{eq} d\tau,
\label{eqER03}
\end{equation}
where according to
Eq. (\ref{eqR09})
\begin{equation}
\langle v(\tau) v(0)\rangle_{eq}={k_b T\over M} E_{\alpha}\left( -\gamma \tau^{\alpha} \right)
\label{eqVACF}
\end{equation}
is the velocity autocorrelation function. 
We note that relation Eq. (\ref{eqER03}) is valid for stationary
processes \cite{remark4}, while the collision model we investigated
in previous section is non-stationary.

 Now consider the constant force field $F(x)=F$, 
using the fractional Kramers Eq. (\ref{eqKr02}). We can show
that a second generalized Einstein relation 
 between
the drift in the presence of the driving field, $\langle x(t) \rangle_F$,
and the mean square displacement, Eq. 
(\ref{eqKr07}),  in the absence of the field is valid
\begin{equation}
\langle x(t) \rangle_F= F {\langle x^2(t) \rangle_{eq}\over  2 k_b T}.
\label{eqEin}
\end{equation}
This relation suggests that the fractional Kramers equation
is compatible with linear response theory \cite{barkaiE,Berlin}.

\subsection{Force free case 2}

 We shall now use the formal solution Eq.
(\ref{eqwww2}) to find moments $\langle  x^{2n}(t)\rangle_{eq}$
for the case $F(x)=0$. Odd moments are equal zero.
Consider the reduced probability density $W_{eq}(x,t)$
of finding the particle on $x$ at time $t$, defined according to
\begin{equation}
W_{eq}(x,t)=\int_{-\infty}^{\infty} dv \int_{-\infty}^{\infty} dv_0 P(x,v,t) M(v_0)
\label{eqFFC01}
\end{equation}
$P(x,v,t)$ is the solution of the fractional Kramers equation
with initial conditions concentrated on $x_0$ and $v_0$.
$M(v_0)$ is Maxwell's probability density implying an equilibrium
initial condition for the initial velocity $v_0$. 

For the standard 
case $\alpha = 1$, $W_{eq}(x,t)$
is a Gaussian, therefore 
$$ \langle x^{2 n } \left( t \right) \rangle_{eq} =
\int_{-\infty}^{\infty} x^{2n} W_{eq}\left( x , t \right) dx=$$
\begin{equation}
 {\left( 2 n \right)! \over 2^n n!} \langle x^2(t) \rangle^n_{eq}
\label{eqfff01}
\end{equation}
and according to \cite{Risken}
\begin{equation}
\langle x^2\left( t \right) \rangle_{eq}=2 {k_b T \over M} 
{ \left[ \gamma_1 t - 1 + \exp\left( - \gamma_1 t \right) \right]
\over
\gamma_1^2}
\label{eqfff02}
\end{equation}
which in the long time limit gives (only for $\alpha = 1$)
\begin{equation}
\langle x^2\left( t \right) \rangle_{eq}\sim 2 {k_b T \over M \gamma_1} t.
\label{eqfff03}
\end{equation}

 According to Eqs. (\ref{eqwww}-\ref{eqwww2}) the calculation of
$\langle x^{2 n } \left( t \right) \rangle_{eq}$, for $0< \alpha< 1$,
follows three steps. First find the Laplace transform of the Gaussian
moments,
$\langle x^{2 n } \left( t \right) \rangle_{eq}$
Eq.
(\ref{eqfff01}).
Since we shall be interested in the long time behavior of
$\langle x^{2 n } \left( t \right) \rangle_{eq}$
it is sufficient to consider only the long time
behavior of the standard $\alpha=1$ case, namely
we use the asymptotic Eq. (\ref{eqfff03}) instead of
Eq. (\ref{eqfff02}). We note that the inclusion of the short
time behavior is also straightforward, but of less interest to us here.
It is easy to show that 
\begin{equation}
\langle x^{2 n } \left( u \right) \rangle_{eq}\sim \left( 2 n \right)! 
\left({k_b T \over M \gamma_1}\right)^n { 1 \over u^{n + 1 } }
\label{eqfff04}
\end{equation}
valid for small $u$ and $\alpha=1$. The second step is to 
 transform
Eq. (\ref{eqfff04}) using
Eq. (\ref{eqwww}) and 
the last step is to invert Laplace transform the result from the second step,
as in
Eq. (\ref{eqwww2}). We find
\begin{equation}
\langle x^{2n} \left( t \right) \rangle_{eq} \sim \left( 2 n \right)! 
\left({ k_b T \over M \gamma_{\alpha}}\right)^n 
{ t^{n(2 - \alpha)} \over \Gamma\left( 2 n - n \alpha +1\right)}.
\label{eqfff05}
\end{equation}
It is easy to check that the moments in (\ref{eqfff05}) can
also be calculated  based on
\begin{equation}
\langle x^{2n} \left( t \right) \rangle \sim {\cal L}^{-1} \left[ \left( { d \over i dk}\right)^{2n}
\left({ u^{1 - \alpha} \over u^{2 - \alpha} + {k_b T \over M \gamma_{\alpha}} k^2}\right)
\left|_{k=0} \right. \right] .
\label{eqfff06}
\end{equation}
This result is also to be expected. 
In  Fourier--Laplace space the well known $\alpha=1$ solution
is
\begin{equation}
W_{eq}\left( k, u \right) \sim {1 \over u + {k_b T \over M \gamma_1} k^2},
\label{eqFFC02}
\end{equation}
if we
use the transformation 
(\ref{eqwww}) in Eq. (\ref{eqFFC02}) we find
\begin{equation}
W_{eq}\left( k, u \right) \sim {u^{1-\alpha} \over u^{2 - \alpha} + {k_b T \over M \gamma_\alpha} k^2}.
\label{eqFFC03}
\end{equation}
And this is the moment generating function in Eq. (\ref{eqfff06}).
We note that in general  there is no guarantee that 
the transformation Eq. (\ref{eqwww}) can be made after
the small $u$ limit of the $\alpha=1$ solution is taken,
instead the small $u$ limit must be taken only after the transformation
Eq. (\ref{eqwww}) takes place.
 The small $(k,u)$ expansion a particular
 coupled CTRW also known as L\'evy walks
has the
same form as Eq. (\ref{eqFFC03}) \{see Eq. $38$ in \cite{KBS}\}
.

 According to Eq. (\ref{eqFFC03}) $W_{eq}(x,t)$ satisfies the following
fractional diffusion equation
\begin{equation}
{\partial W_{eq}\left(x,t\right) \over  \partial t} = 
{k_b T \over M \gamma_\alpha}\  _0 D_t^{-(1 - \alpha)}
{\partial^2 \over \partial x^2} W_{eq}(x,t)
\label{eqFFC04a}
\end{equation}
which is expected to work well only at large  times.
Eq. (\ref{eqFFC04a}) was investigated by Schneider and Wyss \cite{schneider}.

For one dimension the inverse Laplace--Fourier transform of Eq.
(\ref{eqFFC03}) is
\begin{equation}
W_{eq}\left(x,t\right)\sim  
{ \sqrt{\gamma_{\alpha} M} \over \sqrt{2 k_b T}\left(1 - \alpha/2\right)} 
\left( {z^{2 - \alpha/2} \over t^{ 1 - \alpha/2} } \right) 
l_{1 - \alpha/2}\left( z \right)
\label{eqFFC04}
\end{equation}
with 
\begin{equation}
z=t\left[{ \sqrt{k_b T } \over 
\left( \sqrt{2 M \gamma_{\alpha}}  |x|\right) } 
\right]^{{1 \over 1 - \alpha/2}}
\label{eqFFC05}
\end{equation}
The inversion of the two and three dimensional solutions
is not as straightforward as for the one dimensional case, and 
we leave the details for a future publication.

\subsection{Collision Model}

 We have derived the fractional Fokker--Planck equation 
(\ref{eqR04})
from a non-stationary stochastic collision model in the limit of $\epsilon = m/M\to 0$.
Eq. (\ref{eqR04}) describing a fractional Ornstein--Uhlenbeck process,
is the basis of the fractional Kramers equation which is reached
after adding the standard streaming terms (i.e. Newton's law
of motion).
In \cite{barkai3,barkai2} 
the mean square displacement, $\langle x^2(t) \rangle_{col}$ of the test particle was calculated
also
for mass ratios $\epsilon$ not tending to zero. 
It was shown the collision process gives $\langle x^2 \rangle_{col} \sim t^2$
for $0<\alpha< 1$ and finite $\epsilon$, a behavior different from
the asymptotic behavior we have found in Eq.
(\ref{eqKr09})
 based on  the fractional Kramers equation,
$\langle x^2 \rangle \sim t^{ 2 - \alpha}$. The $\langle x^2 \rangle_{col}  \sim t^2$
behavior of the collision model can be understood by the fact
that within the collision process,  one typically finds long 
(collision-free) time intervals,
of the order of the observation time $t$, in which the motion
is ballistic so that $x^2 \sim t^2$. A similar ballistic behavior
is known from the L\'evy walk model with diverging time intervals 
between turning points.
As $\epsilon$ becomes smaller,
the correction to the ballistic term becomes  important
for longer times and these correction terms behave 
like $\langle x^2 \rangle \sim t^{ 2 - \alpha}$.
Also it is clear that the two limits of  $t \to \infty$ and $\epsilon \to 0$  do not commute.
If we first take $t\to \infty$  and only then $\epsilon \to 0$ we find
\begin{equation}
\langle x^2 (t) \rangle_{col} \sim t^2. 
\label{eqcol2}
\end{equation}
Hence the derivation of the fractional Kramers equation
from the stochastic collision model is a delicate matter, with a result
depending on the order in which limits are taken. Another difficulty in our
derivation is that our starting point is a non-stationary process. It is
still unclear if stationary models can lead to dynamics
described by the fractional Kramers equation.


%
\section{Summary and Discussion}
\label{secSum}

 We have investigated a fractional Kramers equation
which has the following properties:\\
{\bf (a)} the velocity of the particle evolves according
to a
fractional Ornstein--Uhlenbeck process described by Eq. 
(\ref{eqR04}),
the velocity moments decay according to a  Mittag--Leffler relaxation, 
namely as a stretched exponential (Kohlrausch form) for short times and as
a power law for long times,\\
{\bf (b)} in the absence of a force field diffusion is enhanced, $1< \delta < 2$,\\
{\bf (c)} the stationary solution of the fractional Kramers equation
 is the Maxwell-Boltzmann distribution,\\
{\bf (d)} Einstein relations are obeyed in consistency
with fluctuation dissipation theorem and linear response theory,\\
{\bf (e)} in Laplace space a simple transformation of solutions
of ordinary Kramers equation gives the solution of
fractional Kramers equation.\\

 As mentioned in the
introduction fractional kinetic equations in the literature
are related to the CTRW.
We showed here that in the small $(k,u)$ limit
$W_{eq}(k,u)$ has the same form as a particular coupled Levy walk CTRW.
Other limits of the CTRW are shown to correspond to
other fractional kinetic equations. 
As pointed out in \cite{compte}
the fractional diffusion equation \cite{schneider}, describing the sub-diffusion, corresponds
to the uncoupled CTRW in the limit of small $(k,u)$.
A comparison between the uncoupled CTRW and solution of fractional
diffusion equation in $(x, t )$ space was carried out in \cite{barkai9}.
Another fractional
equation in  
\cite{fogedby,zaslavsky}
describes L\'evy flights for which $\langle x^2 \rangle = \infty$,
such a fractional equation is related to the decoupled limit of 
the CTRW with diverging jump
lengths.

 One may ask if
it is worthwhile to introduce fractional
derivatives, given that the older CTRW approach is so successful.
Besides the fact that fractional equations are beautiful and 
simple (i.e., in some cases
they are solvable),
these equations can incorporate the
effect of an external potential field. To us, this extension
seems important although  not explored in depth in the present paper.
Little is known on anomalous diffusion in an external force field.

 Metzler, Barkai and Klafter \cite{MBK} 
have investigated a fractional Fokker--Planck 
equation defined with the fractional Liouville
--Riemann operator. In the absence of an
external force  the fractional Fokker--Planck equation
investigated in \cite{MBK} describes a sub--diffusive behavior
($\delta < 1$). The equation considers a type of over damped dynamics
in which only the coordinate $x$ is considered not the velocity $v$.
The fractional Fokker--Planck equation in \cite{MBK} together with
the fractional Kramers equation investigated here give a stochastic 
framework for both sub and enhanced
diffusion in an external field. We believe that both approaches
will find their application.


{\em Note added in proof.} Recently related 
work on fractional diffusion was published \cite{add}.

\section{Acknowledgments}

EB thanks J. Klafter and R. Metzler for helpful discussions. 
 This research was 
supported in part by a grant from the NSF.

\section{Appendix A}

 We find the  Laplace transform of 
Eq. (\ref{eqlevy}) 
\begin{equation}
 R_{s}( u ) = \int_0^{\infty} e^{ - u t} {1 \over  \alpha \gamma_{\alpha} t^{\alpha} } z^{ \alpha + 1} l_{\alpha} \left( z \right) dt,
\label{eqAn01}
\end{equation}
with $z$ defined in Eq. (\ref{eqlevyz}).
Using the change of variables
$t=y (s/\gamma_{\alpha})^{1 /\alpha}$ it is easy to show
$(s \ne 0)$
$$ R_{s}\left( u \right) = -{ 1 \over \alpha s} {d \over du}\int_0^{\infty}
\exp\left[ - y \left({ s\over \gamma_{\alpha} } \right)^{ 1 / \alpha} u \right]
l_{\alpha}\left( y \right) dy =$$
$$ -\left({ 1 \over \alpha s} \right) {d \over du} \exp\left( - {  s \over \gamma_{\alpha}} u^{\alpha} \right) = $$
\begin{equation}
{u^{\alpha - 1} \over \gamma_{\alpha} } \exp\left( - { s u^{\alpha} \over \gamma_{\alpha} } \right),
\label{eqAn02}
\end{equation}
which is  $R_{s}( u )$, Eq.  
(\ref{eqA11m}). 
From Eq. (\ref{eqA09}) we learn that Eq. (\ref{eqA11m}) is valid also for $s=0$.

\newpage

{\bf Figure Caption}

Figure 1: The dynamics of $Q(v,t)$ for the fractional Ornstein-Uhlenbeck
process with $\alpha = 1/2$ and for times, t = 0.02,0.2,2, 20 
(solid, dashed, dotted, and dot-dash lines, respectively). Also shown (fine
dotted curve) is the stationary solution which is Maxwell's distribution. 
Notice the cusp on $v = v_0 = 1$ as well as the non-symmetrical shape 
of $Q(v,t)$. 


\begin{thebibliography}{99}

\bibitem{SM} H. Scher and E. Montroll, Phys. Rev. {\bf B12},2455 (1975) 

\bibitem{SL} H. Scher and M. Lax, Phys. Rev. {\bf B7}, 4491 (1973); Phys. Rev
{\bf B7}, 4502 (1973)

\bibitem{Weiss1} G. H. Weiss {\em  Aspects and Applications of the Random Walk}
               North Holland (Amsterdam -- New York -- Oxford, 1994)

\bibitem{Bouch} J.--P. Bouchaud and A. Georges, Phys. Rep.
{\bf 195}, 127 (1990)

%
\bibitem{Klafter1} J. Klafter, M. F. Shlesinger and G. Zumofen, {\em Phys. Today
} {\bf 49}, 33 (1996).

\bibitem{Balescu} R. Balescu, {\it Statistical Dynamics; Matter Out of Equilibrium} Imperial College Press, World scientific, Singapore (1997).

\bibitem{barkai8} E. Barkai and J. Klafter, 
Lecture Notes in Physics, S. Benkadda and G. M. Zaslavsky
Ed. Chaos, Kinetics and Non-linear Dynamics in Fluids and Plasmas
(Springer-Verlag, Berlin 1998).

\bibitem{schneider} W. R. Schneider and  W. Wyss, J. Math. Phys.
{\bf 30}, 134 (1989)

\bibitem{Gl} W. G. Gl$\ddot{o}$ckle and T. F. Nonnenmacher, Macromolecules, {\bf 24}
6426 (1991)

\bibitem{fogedby} H. C. Fogedby, Phys. Rev. Lett. {\bf 73}, 2517 (1994);
Phys. Rev. E {\bf 58}, 1690 (1998);S. Jespersen, R. Metzler and H. C. Fogedby, Phys. Rev. E {\bf 59},
2736 (1999)


\bibitem{zaslavsky}
G. M. Zaslavsky, M. Edelman and B. A. Niyazov,
Chaos {\bf 7}, 159 (1997)

\bibitem{saichev} A. I. Saichev and M. Zaslavsky, Chaos {\bf 7} 4 (1997)


\bibitem{MBK} R. Metzler, E. Barkai and J. Klafter, Phys. Rev. Lett. {\bf 82},
             3563 (1999) 


\bibitem{Hui} T. Huillet, J. Phys. A  {\bf 32} 7225 (1999)

\bibitem{Grig} Grigolini P, Rocco A, West B. J., Phys. Rev. E. 
{\bf 59} 2603 (1999)

\bibitem{kuz} D. Kusnezov. A. Bulgac and G. Do Dang,
Phys. Rev. Lett. {\bf 82}, 1136 (1999)

\bibitem{Zum}  G. Zumofen and J. Klafter
{\em Chem. Phys. Lett.}  {\bf 219} 303 (1994) 

\bibitem{SB} E. Barkai and R. Silbey, Chem. Phys. Lett., {\bf 310} 287 (1999).

\bibitem{MUK} V. Chernyak, M. Schultz and S. Mukamel, J. Chemical Physics,
{\bf 111} 7416 (1999)

\bibitem{oldham} K. B. Oldham and J. Spanier, {\em The
Fractional Calculus} Academic Press, (New York) 1974.

\bibitem{Hilfer} R. Hilfer and L. Anton, Phys. Rev. E. {\bf 51},
R848 (1995)
  

\bibitem{compte} A. Compte,  Phys. Rev. E {\bf   53},  4191 (1996)

\bibitem{West} B. J. West, P. Grigolini, R. Metzler and T. F. Nonnenmacher,
Phys. Rev. E {\bf 55}, 99 (1997)


\bibitem{barkai9} E. Barkai, R. Metzler and J. Klafter,  Phys. Rev. E.
Phys. Rev. E {\bf 61} 132 (2000)


%

\bibitem{Ral} Lord Rayleigh {\em Phil. Mag.} {\bf 32} 424 (1891). [{\em Scientific
Papers} (Cambridge 1902) {\bf 3}, 473]

%
\bibitem{Kampen} N.G. van Kampen {\em Stochastic Processes in Physics and
               Chemistry} North Holland (Amsterdam -- New York -- Oxford, 1981)



\bibitem{barkai3}  E. Barkai and V. N.  Fleurov, {\em J. Chem. Phys.} {\bf 212},
             69, (1996) 

\bibitem{barkai2}  E. Barkai and V. N.  Fleurov, {\em Phys. Rev. E.} {\bf 56},
             6355, (1997) 

\bibitem{MKp} R. Metzler and J. Klafter (preprint)

\bibitem{Risken} H. Risken, {\it The Fokker--Planck
equation\/} (Springer, Berlin, 1989)


\bibitem{remark3} To see this Laplace transform  Eq. (\ref{eqR04}) using Eq.
(\ref{eqL05})
%
$$ u Q(u,v) - Q(v,t=0) = $$
$$\gamma_{\alpha} \hat{L}_{fp}
\left[ u^{1 - \alpha} Q(v,u) - _0D_t^{ - \alpha} Q(v,t) |_{t = 0} \right], $$
setting $_0D_t^{ - \alpha} Q(v,t) |_{t = 0}=0$ 
we get an expression equivalent to
to the Laplace transform of
Eq. (\ref{eqR03}).

\bibitem{Feller} W. Feller,{\em  An introduction to probability Theory and
Its Applications} Vol. 2 (John Wiley and Sons, New York 1970).


%








\bibitem{Zol} V. M. Zoltarev, Dokl. Acad. Nauk. USSR {\bf 98}, 715 (1954)

\bibitem{EWM} E. W. Montroll and B. J. West in {\it Fluctuation Phenomena} Eds. E. W. Montroll
and J. L. Lebowitz (North Holland, Amsterdam, 1987)

\bibitem{Hangi} P. H$\ddot{a}$nggi, P. Talkner and M. Borkovec, Rev. Mod. Phys.
{\bf 62}, 251 (1990)

\bibitem{Mont} J. A. Montgomery, D. Chandler, and B. J. Berne,
J. Chem. Phys. {\bf 70}, 4056 (1979)

\bibitem{Knessel} C. Knessel, M. Mangel, B. J. Matkowsky, A. Schuss,
and C. Tier,  J. Chem. Phys. {\bf 81} 1285 (1984)

\bibitem{Bork} M. Borkovek, J. E. Straub, and B. J. Berne,
J. Chem. Phys. {\bf 85}, 146 (1986)


\bibitem{barkaif} E. Barkai and V. Fleurov Phys. Rev. E {\bf 52} 1558
(1995)

\bibitem{Ber1} D. J. Bicout, A. M. Berezhkovskii, A. Szabo and G. H. Weiss, 
Phys. Rev. E. {\bf 59} 3702 (1999) 

\bibitem{Ber2} A. M. Berezhkovskii, D. J. Bicout and G. H. Weiss,
J. Chem. Phys. {\bf 111} 11050 (1999) 

\bibitem{Erde} A. Erdelyi Ed. {\it  Higher Transcendental Functions} McGraw-Hill, New York (1955) 


\bibitem{remark4}
To see this write
$$\langle x^2 (t) \rangle_{eq}= \langle \left[\int_0^t v\left(t_1\right) d t_1\right]
\left[\int_0^t v\left(t_2\right) d t_2\right] \rangle_{eq},$$
%
if the process is  stationary 
$\langle v(t_1) v(t_2)\rangle_{eq} = \langle v(\tau) v(0)\rangle_{eq}$
%
with $\tau=|t_1 - t_2|$, and only then  Eq. (\ref{eqER03}) valid.

\bibitem{barkaiE} 
E. Barkai and V. N. Fleurov, Phys. Rev. E {\bf 58},
1296 (1998)

\bibitem{Berlin} 
Y. A. Berlin, L. D. A. Siebbeles and A. A. Zharikov,
Chem. Phys. Lett. {\bf 305} 123 (1999)

\bibitem{KBS} 
J. Klafter, A. Blumen and M. F. Shlesinger, 
Phys. Rev. A {\bf 35}, 3081 (1987)


%

%
\bibitem{add}
M. Bologna, P. Grigolini P  and J.  Riccardi,
Phys. Rev. E {\bf 60} 6435 (1999).
R. Kutner, K. Wysocki
{\em Physica A} {\bf 274} 67 (1999).
V. V. Yanovsky, A. V. Chechkin, D. Schertzer
and A. V. Tour nlin/001035 (to appear in Physica A)




\end{thebibliography}
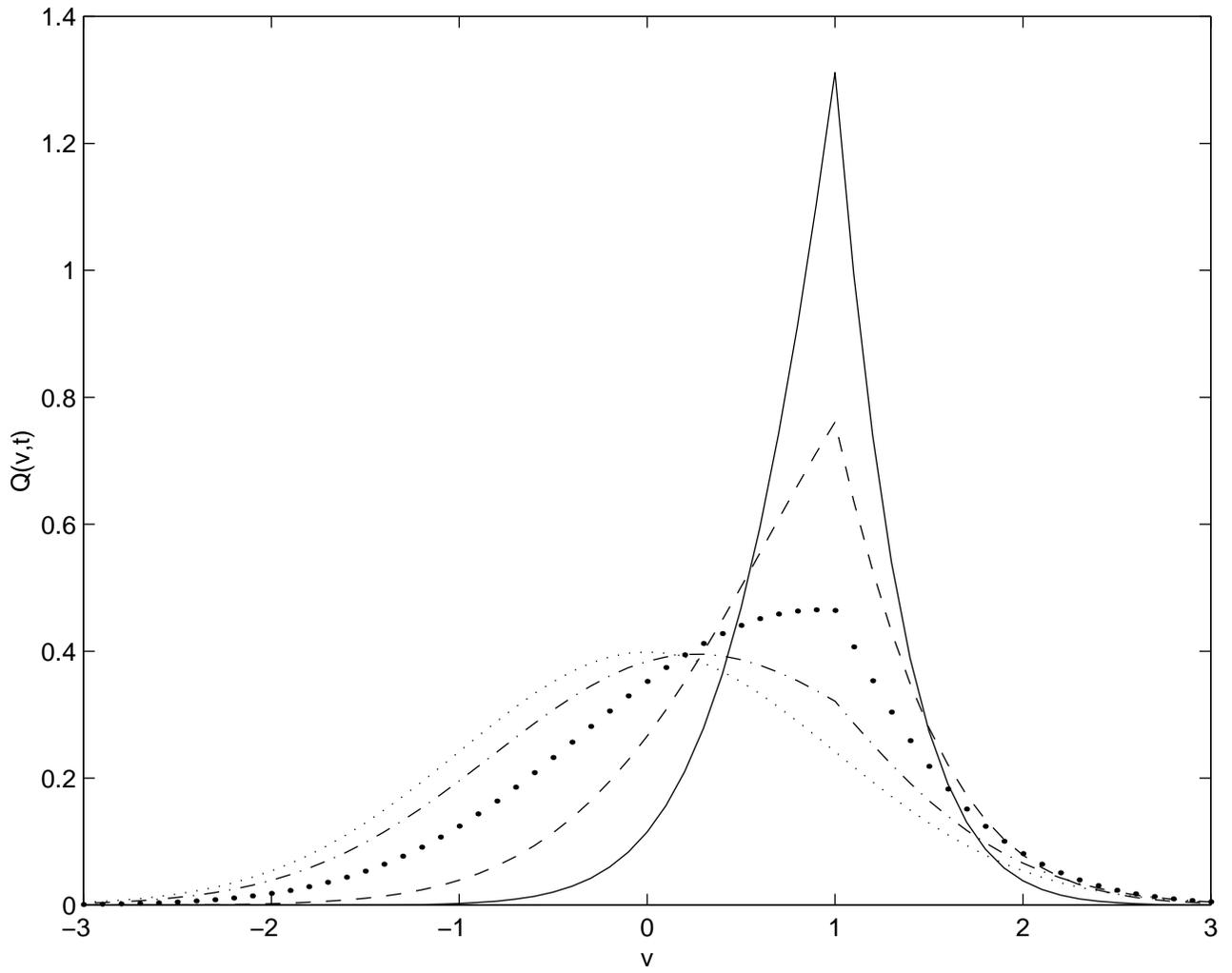
\end{document}